\documentclass[aps,showpacs,superscriptaddress,groupedaddress]{revtex4_1}
\usepackage{graphicx}
\usepackage{amsmath}
\usepackage{bm}
\def\supit#1{\raisebox{0.8ex}{\small\it #1}\hspace{0.05em}}  
\def\skiplinehalf{\medskip\\}
\draft % marks overfull lines with a black rule on the right
\begin{document}

%\preprint{}

\title{POLARBEAR-2: an instrument for CMB polarization measurements} %Title of paper


\author{Y. Inoue\supit{a,b}, P. Ade\supit{c}, Y. Akiba\supit{d}, C. Aleman\supit{e,ab} , K. Arnold\supit{f}, C. Baccigalupi\supit{g}, B. Barch\supit{h}, D. Barron\supit{h}, A. Bender\supit{i,z}, D. Boettger\supit{j}, J. Borrill\supit{k,y}, S. Chapman\supit{l}, Y. Chinone\supit{h}, A. Cukierman\supit{h}, T. de Haan\supit{h}, M. A. Dobbs\supit{m}, A. Ducout\supit{n}, R. Dunner\supit{j}, T. Elleflot\supit{e,ab}, J. Errard\supit{o}, G. Fabbian\supit{g}, S. Feeney\supit{n}, C. Feng\supit{p}, G. Fuller\supit{e,ab}, A. J. Gilbert\supit{m}, N. Goeckner-Wald\supit{h}, J. Groh\supit{h}, G. Hall\supit{h}, N. Halverson\supit{q,aa,ac}, T. Hamada\supit{b}, M. Hasegawa\supit{b,d}, K. Hattori\supit{b}, M. Hazumi\supit{b,d,r,v}, C. Hill\supit{h}, W. L. Holzapfel\supit{h}, Y. Hori\supit{h}, L. Howe\supit{e,ab}, F. Irie\supit{r,s}, G. Jaehnig\supit{q,ac}, A. Jaffe\supit{n}, O. Jeong\supit{h}, N. Katayama\supit{r,s}, J. P. Kaufman\supit{e,ab}, K. Kazemzadeh\supit{a,ab}, B. G. Keating\supit{e,ab}, Z. Kermish\supit{t}, R. Keskital\supit{h,k}, T. Kisner\supit{k,y}, A. Kusaka\supit{k}, M. Le Jeune\supit{u}, A. T. Lee\supit{h,k}, D. Leon\supit{e,ab}, E. V. Linder\supit{k}, L. Lowry\supit{e,ab}, F. Matsuda\supit{e,ab}, T. Matsumura\supit{v}, N. Miller\supit{w}, K. Mizukami\supit{r,s}, J. Montgomery\supit{m}, M. Navaroli\supit{e,ab}, H. Nishino\supit{b}, H. Paar\supit{e,ab}, J. Peloton\supit{u}, D. Poletti\supit{u}, G. Puglisi\supit{g}, C. R. Raum\supit{h}, G. M. Rebeiz\supit{ab}, C. L. Reichardt\supit{y}, P. L. Richards\supit{h}, C. Ross\supit{l}, K. M. Rotermund\supit{l}, Y. Segawa\supit{d}, B. D. Sherwin\supit{h,ad}, I. Shirley\supit{h}, P. Siritanasak\supit{e,ab}, N. Stebor\supit{e,ab}, R. Stompor\supit{u} A. Suzuki\supit{h,y}, O. Tajima\supit{b,d}, S. Takada\supit{x}, S. Takatori\supit{d}, G. P. Teply\supit{e,ab}, A. Tikhomirov\supit{l}, T. Tomaru\supit{b}, N. Whitehorn\supit{h}, A. Zahn\supit{e}, and O. Zahn\supit{h}
\skiplinehalf
\supit{a}Institute of physics, Academia Sinica, Taipei, Taiwan(R.O.C.); \\
\supit{b}High energy accelerator research organization, Tsukuba, Japan; \\
\supit{c}School of Physics and Astronomy, Cardiff University, Cardiff CF10, 3XQ, UK;\\
\supit{d}SOKENDAI, The Graduate Institute for Advanced Studies, Hayama, Miura District, Kanagawa 240-0115, Japan; \\
\supit{e}Center for Astrophysics and Space Science, University of California, San Diego, CA, 92093, USA; \\
\supit{f}Department of Physics, University of Wisconsin, Madison, WI, 53706, USA; \\
\supit{g}Scuola Internazionale Superiore di Studi Avanzati (SISSA), Via Bonomea 265, 34136, Trieste, Italy; \\ 
\supit{h}Department of Physics, University of California, Berkeley, CA, 94720, USA; \\
\supit{i}Argonne National Laboratory, Argonne, IL 60439, USA; \\
\supit{j}Department of Astronomy, Pontifica University Catolica de Chile, Santiago, Chile; \\
\supit{k}Computational Cosmology Center, Lawrence Berkeley National Laboratory, Berkeley, CA 94720 USA; \\
\supit{l}Department of Physics and Atmospheric Science, Dalhousie University, Halifax, NS, B3H 4R2, Canada; \\
\supit{m}Physics Department, McGill University, Montreal, QC, H3A 0G4, Canada; \\
\supit{n}Department of Physics, Blackett Laboratory, Imperial College London, London SW7 2AZ UK; \\
\supit{o}Sorbonne Universites, Institut Lagrange de Paris (ILP), 75014 Paris, France; \\
\supit{p}Department of Physics and Astronomy, University of California, Irvine, CA, 92697, USA; \\
\supit{q}Center for Astrophysics and Space Astronomy, University of Colorado, Boulder, CO, 80309, USA; \\
\supit{r}Kavli IPMU (WPI), UTIAS, The University of Tokyo, Kashiwa, Chiba 277-8583, Japan; \\
\supit{s}Yokohama National University, Yokohama, Japan; \\
\supit{t}Department of Physics, Princeton University, Princeton, NJ, 08544, USA; \\
\supit{u}AstroParticule et Cosmologie, Univ Paris Diderot, CNRS/IN2P3, CEA/Irfu, Obs de Paris, Sorbonne, Paris Cite, France; \\
\supit{v}Institute of Space and Astronautical Studies (ISAS), Japanese Aerospace Exploration Agency (JAXA), Sahamihara, Kanagawa 252-510, Japan; \\
\supit{w}Observational Cosmology Laboratory, Code 665, NASA Godard Space Flight Center, Greenbelt, MD, 20771, USA; \\
\supit{x}National Institute for Fusion Science 322-6 Oroshi-cho, Toki City, GIFU Prefecture, Japan; \\
\supit{y}Space Sciences Laboratory, University of California, Berkeley, CA, 94720, USA;\\
\supit{z}Department of Astronomy and Astrophysics, University of Chicago, Chicago, IL 60637, USA;\\
\supit{aa}Department of Astrophysical and Planetary Sciences, University of Colorado, Boulder, CO, 80309, USA;\\
\supit{ab}Department of Electrical and Computer Engineering, University of California, San Diego, CA, 92093, USA;\\
\supit{ac}Department of Physics, University of Colorado, Boulder, CO, 80309, USA;\\
\supit{ad}Miller Institute for Basic Research in Science, University of California, Berkeley, CA, 94720, USA;
}

\begin{abstract}
POLARBEAR-2 (PB-2) is a cosmic microwave background (CMB) polarization experiment that will be located in the Atacama highland in Chile at an altitude of 5200~m.
Its science goals are to measure the CMB polarization signals originating from both
primordial gravitational waves and weak lensing.
PB-2 is designed to measure the tensor to scalar ratio, $r$, with precision $\sigma(r) < 0.01$, and the sum of neutrino masses, $\Sigma m_{\nu}$ , with $\sigma(\Sigma m_{\nu} ) < 90~\mathrm{meV}$. 
%$The PB-2 receiver system is designed to achieve the sensitivity of tensor to scalar ratio, $r\sim0.01$ (95\% C.L.), 
%and the sum of neutrino masses, $\Sigma m_{\nu}$=90~meV (68 \% C.L.). 
To achieve these goals, PB-2 will employ 7588 transition-edge sensor bolometers at 95~GHz and 150~GHz, 
which will be operated at the base temperature of 270 mK.
Science observations will begin in 2017.     
\end{abstract}

\maketitle %% required
\noindent
\section{Introduction}
\label{sec:intro}  % \label{} allows reference to this section
\begin{figure}
   \begin{center}
   \begin{tabular}{c}
   \includegraphics[height=8cm]{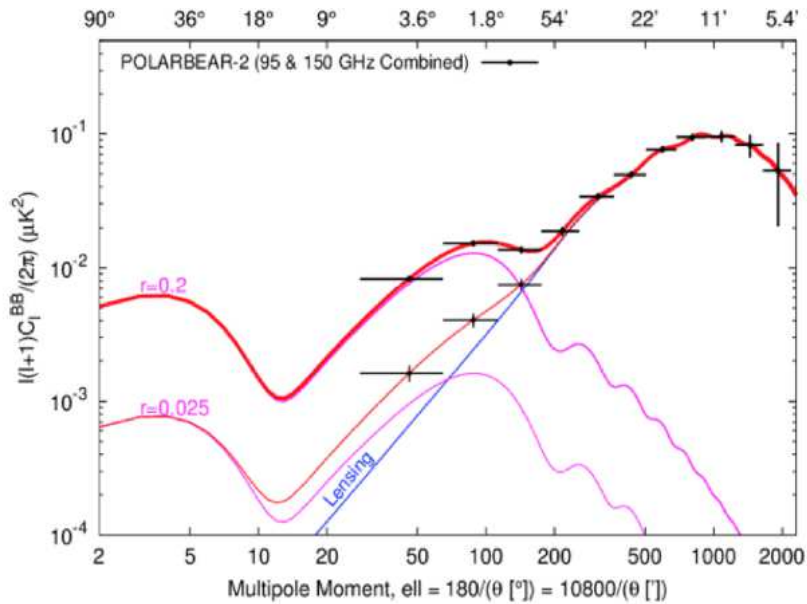}
   \end{tabular}
   \end{center}
   \caption[example] 
%>>>> use \label inside caption to get Fig. number with \ref{}
   { \label{fig:SEN} 
Expected sensitivity of the PB-2 receiver for B-mode detection within 1 sigma errors for 3 years of observation. Pink and blue curves are the power spectra for primordial gravitational wave and gravitational lensing B-modes, respectively.}
   \end{figure}

PB-2 is a ground-based CMB polarization experiment with a large detector array that consists of 7588 dual-band antenna-coupled Al-Mn transition edge sensor (TES) bolometers for simultaneous measurements at 95 and 150 GHz~\cite{Nate2016,Takayuki2012}. The main goal of PB-2 is to detect degree scale odd-parity (B-mode) polarization patterns~\cite{Kamionkowski}. The B-mode is created by primordial gravitational waves generated during the inflation~\cite{Sato:1980yn,Guth:1980zm}. It is a smoking gun signature of inflationary universe. 
PB-2 is designed to measure the inflation models corresponding to the tensor to scalar ratio, $r$, with precision $\sigma(r) < 0.01$.
PB-2 also plans to measure the sub-degree scale B-mode from gravitational lensing, whose amplitude is sensitive to the sum of neutrino masses, $\Sigma m_{\nu}$ , with $\sigma(\Sigma m_{\nu} ) < 90~\mathrm{meV}$. 
The expected sensitivity is shown in Fig.~\ref{fig:SEN}.
We plan to start the scientific observation from early 2017.

%%%%%%%%%%%%%%%%%%%%%%%%%%%%%%%%%%%%%%%%%%%%%%%%%%%%%%%%%%%%%
\section{Instrument overview} 
\begin{figure}
   \begin{center}
   \begin{tabular}{c}
   \includegraphics[height=10cm]{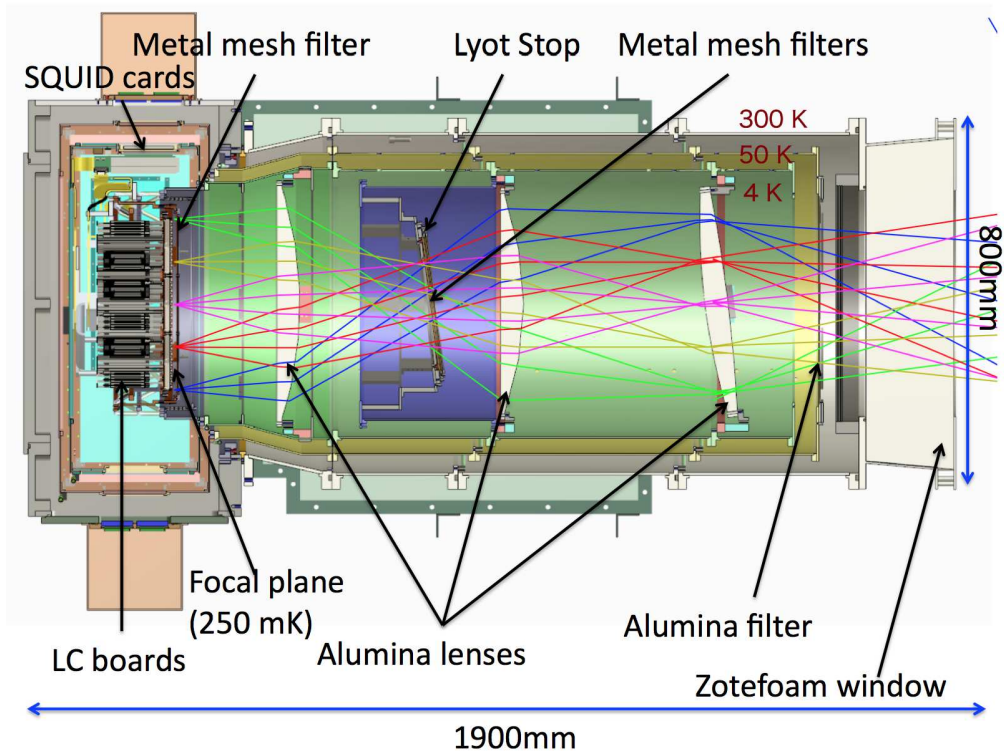}
   \end{tabular}
   \end{center}
   \caption[example] 
%>>>> use \label inside caption to get Fig. number with \ref{}
   { \label{fig:RE} 
Cross section of the PB-2 receiver system.}
   \end{figure}

The drawing of the PB-2 receiver system is shown in Fig.~\ref{fig:RE}. It consists of the box-type cryostat for the focal plane, and the optics tube for lenses and filters~\cite{Yuki2014,Yuki1}.
 The PB-2 receiver will be housed in a new telescope with the POLARBEAR-1 (PB-1) design~\cite{Nate2016}. The detector and refractive optical system are cooled with a combination of two pulse tube coolers and a sorption cooler~\cite{Takayuki2012,PT415,CRcryo}. The measured temperature and holding time of focal plane are 270 mK and 28 hours~\cite{Yuki_thesis}. 
The target noise equivalent temperature~(NET) of each detector is $360~\mathrm{\mu K \sqrt{s}}$. Total array NET of each frequency is $5.8~\mathrm{\mu K \sqrt{s}}$ and the combined array NET is $4.1~\mathrm{\mu K \sqrt{s}}$. The specifications of PB-2 system are shown in Table.~\ref{tab:spec}. We mount the polarization modulator and gain calibrator into the telescope.
We plan to use the sapphire half wave plate at the front of the vacuum window as the polarization modulator~\cite{Charlie2016}.
This modulator allows to reduce the $1/f$ noise and mitigate the beam systematics.
The operation frequency is 2~Hz. The signal frequency is 8 Hz due to 4f modulation with birefringent material.
We also place the chopped thermal source at the backside of secondary mirror as a gain calibrator. For the gain calibration we employ the non-polarized thermal source at 1000 K.
The chopped frequencies are between 5 and 80 Hz.

\begin{table}[h]
\caption{The summary of the PB-2 receiver specifications.} 
\label{tab:spec}
\begin{center}       
\begin{tabular}{| l | c |} %% this creates two columns
%% |l|l| to left justify each column entry
%% |c|c| to center each column entry
%% use of \rule[]{}{} below opens up each row
\hline
\rule[-1ex]{0pt}{3.5ex}  $ $  &  \textbf{POLARBEAR-2} \\
\hline
\rule[-1ex]{0pt}{3.5ex}  Frequencies & $\mathrm{95\,GHz}$ and $\mathrm{150\,GHz}$  \\
\hline
\rule[-1ex]{0pt}{3.5ex}  Number of pixels &1897 (7588 bolometers) \\
\hline
\rule[-1ex]{0pt}{3.5ex}  NET (bolometer) & $\mathrm{360/360\,\mu K \sqrt{s}}$ ($\mathrm{95/150\,GHz}$) \\
\hline
\rule[-1ex]{0pt}{3.5ex}  NET (array) & $\mathrm{5.8/5.8\,\mu K \sqrt{s}}$ ($\mathrm{95/150\,GHz}$) \\
\rule[-1ex]{0pt}{3.5ex}  $ $  & $\mathrm{4.1\,\mu K \sqrt{s}}$ (95 and 150\,GHz combination) \\
\hline
\rule[-1ex]{0pt}{3.5ex}  Focal plane & $\mathrm{270\,mK}$ \\
 \rule[-1ex]{0pt}{3.5ex}   Temperature   &  \\
\hline
\rule[-1ex]{0pt}{3.5ex}  Field of View & $\mathrm{4.8\, ^\circ}$ \\
\hline
\rule[-1ex]{0pt}{3.5ex}  Beam Size  &5.2\,arcmin. @95\,GHz, 3.5\,arcmin. @150\,GHz \\
\hline
\rule[-1ex]{0pt}{3.5ex}  Sky Coverage& 80\,\% \\
\hline
%\rule[-1ex]{0pt}{3.5ex}  Observation Efficiency & 18\,\% & 18\,\% \\
%\hline

\end{tabular}
\end{center}
\end{table}

%%-----------------------------------------------------------
\section{Optics} 
\label{sec:opt}
\begin{figure}
   \begin{center}
   \begin{tabular}{c}
   \includegraphics[height=9cm]{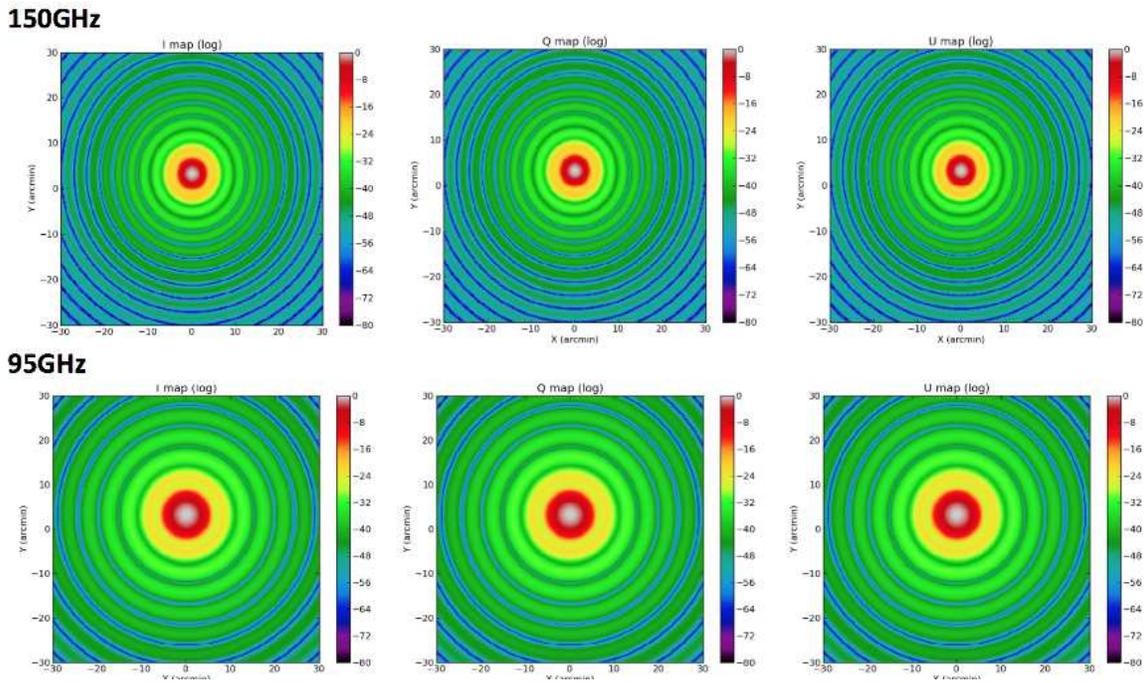}
   \end{tabular}
   \end{center}
   \caption[example] 
%>>>> use \label inside caption to get Fig. number with \ref{}
   { \label{fig:Beam} 
Beam simulation of PB-2 optical system. I (left), Q (middle) and U (right) maps are shown for 150 GHz (top) and 95 GHz (bottom). This simulation includes the reflective alumina lenses and the Gregorian mirror system. We use the QUAST simulation~\cite{QUAST}, which is add-on software of GRASP~\cite{GRASP}. }
   \end{figure}

The PB-2 optics consists of combination of an off-axis Gregorian telescope and alumina re-imaging lenses~\cite{Yuki2014,Yuki_thesis}.
The primary and secondary mirrors meet the Mizuguchi-Dragon condition~\cite{Dragone,Mizuguchi} to cancel the aberration and cross-polarization.
 The size of the primary mirror is 3.5 m in diameter projected along boresight with 2.5 m high-precision monolithic mirror and a 1 m guard ring. Its diameter corresponds to the angular resolution of 3.5 and 5.2 arcmin. at 150 and 95 GHz band, respectively. 
 The secondary mirror is 1.5 m in diameter.
 The simulated beams are shown in Fig.~\ref{fig:Beam}.
 The prime focus baffle is between the primary and secondary mirror to reduce the unexpected stray light.
The alumina re-imaging lenses are with 99.9~\% purity from Nihon ceratech~\cite{NS}
and are placed at 4 K stage in the PB-2 receiver cryostat. The diameters and thicknesses of alumina lenses are 500 mm and 50 mm, respectively.
The positions of the alumina lenses are measured precisely with a laser tracker.
%The estimated Strehl ratio with measured lens positions are shown in Fig.~\ref{}. 
We place a Lyot stop between the aperture and collimator lenses. The Lyot stop reduced the stray light and defines the beam shapes. The edge of Lyot stop is 180 mm in diameter. 
We use new absorber compound named ``KEK Black" as the material of the Lyot stop~\cite{Yuki_thesis}.
%The KEK black is newly invented by elsewhere. 
%The composition of KEK black by mass is: 64 \% Stycast 1090 (Emerson \& Cuming, Inc., Woburn, MA [57]); 6 \% Catalyst 9 [57]; 26 \% Carbon black(MITSUBISHI Carbon Black \#10 [58]); 4 \% Powder Beads (Mogu corporation [59]). 

 \subsection{Vacuum window and infrared filters}
The Zotefoam~\cite{Zote} window is employed as the vacuum window. The refractive index of Zotefoam window is close to unity, so that we can regard it as vacuum in the range of millimeter wavelengths.
The diameter and thickness of window are 800 mm and 200 mm, respectively.
The measured deformation-depth of the vacuum window is 50~mm.
To reduce IR emission from the window, a set of IR filters is placed at each thermal stage.
The RT-MLI~\cite{RT_MLI} as a 300 K filter is placed at the backside of the Zotefoam window.
The cutoff frequency and temperature of RT-MLI is about 400 GHz and 168 K.
We employ the alumina filter as the 50~K absorptive filter.
We newly developed a high-thermal-conductivity infrared (IR) filter using alumina~\cite{Yuki1}. The diameter of alumina filter is 460 mm. We estimated the 3 dB cutoff frequency using a Fourier-transform spectrometer. The 3 dB cutoff frequency is 650 GHz~\cite{Yuki2}. The cut-off shape is steeper than that of conventional filters. The high thermal conductivity of an alumina minimizes thermal gradients. The temperature rise of the alumina filter is only 3 \% of the conventional filter~~\cite{Yuki2014}. 
We also employ the metal mesh filters at 4~K and 350 mK stages. The cutoff frequencies of filters are listed in Table~\ref{tab:filter}.

 \begin{table}[h]
\caption{IR filter specifications. All the temperatures and transmittances are measured values. The 3 dB cutoff of RT-MLI and alumina filter are quoted from elsewhere~\cite{Yuki2,RT_MLI}. The 3 dB cutoff of metal mesh filters are measured by Cardiff group. } 
\begin{center}       
\begin{tabular}{|c|c|c|c|c|} %% this creates two columns
%% |l|l| to left justify each column entry
%% |c|c| to center each column entry
%% use of \rule[]{}{} below opens up each row
\hline
\rule[-1ex]{0pt}{3.5ex}  Filter &Temperature & 3 dB & Transmittance& Transmittance\\
\rule[-1ex]{0pt}{3.5ex}  (stage)& &Cutoff & (95~GHz)& (150~GHz)\\ \hline
\rule[-1ex]{0pt}{3.5ex}  RT-MLI (300 K)&168 K & 2000~GHz & 99 \%& 99 \% \\ \hline
\rule[-1ex]{0pt}{3.5ex}  Alumina filter (50 K) & 55~K & 650 GHz & 96 \%& 96 \% \\ \hline
\rule[-1ex]{0pt}{3.5ex}  Metal mesh filter (4 K)& 5.8 K & 360 GHz & 96 \%& 94 \% \\ \hline
\rule[-1ex]{0pt}{3.5ex}  Metal mesh filter (4 K)& 5.8 K & 261 GHz & 98 \%& 94 \% \\ \hline
\rule[-1ex]{0pt}{3.5ex}  Metal mesh filter (350 mK)& 0.5 K & 171 GHz & 94 \%& 92 \% \\ \hline

%\rule[-1ex]{0pt}{3.5ex}  Observation Efficiency & 18\,\% & 18\,\% \\
%\hline
\end{tabular}
\label{tab:filter}
\end{center}
\end{table} 
 
\subsection{Anti-reflection coating}
The broadband anti-reflection (AR) coating can increase the efficiency of the high-reflection optical elements, such as silicon lenslets, alumina lenses and alumina filters. The typical refractive indices of these elements are $\sim3$ corresponding to 25 \% reflection at the surface.

\begin{figure}
   \begin{center}
   \begin{tabular}{c}
   \includegraphics[height=8cm]{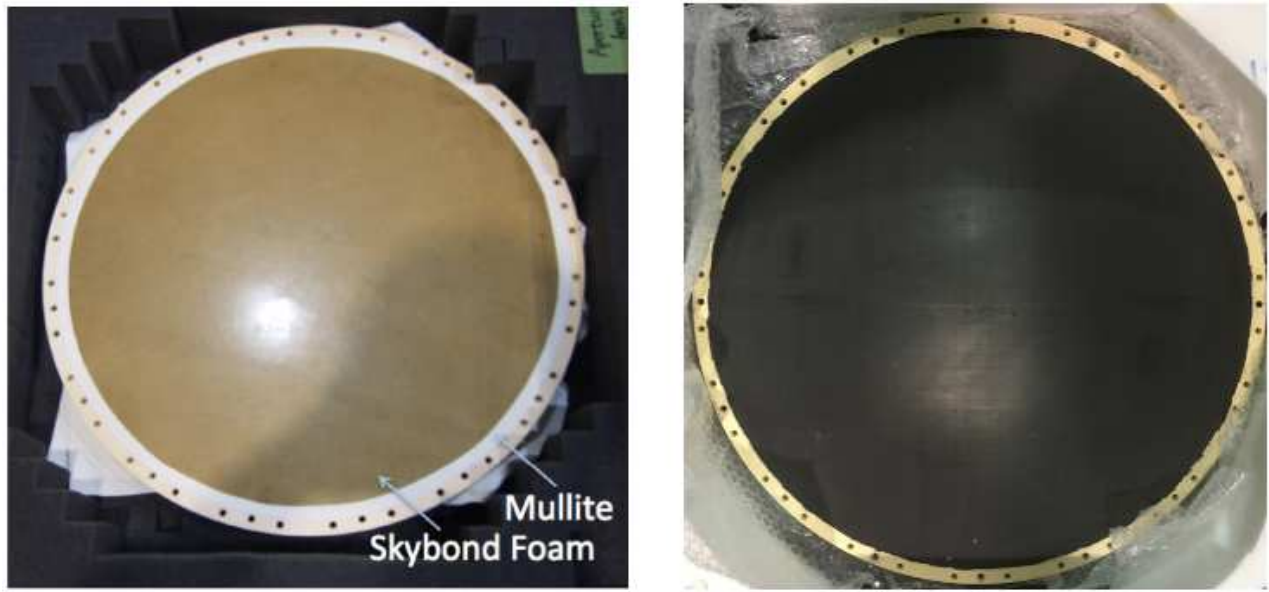}
   \end{tabular}
   \end{center}
   \caption[example] 
%>>>> use \label inside caption to get Fig. number with \ref{}
   { \label{fig:AR} 
   Left: An alumina lens with Skybond$+$mullite AR coating. Right: An alumina lens with epoxy coating. We apply the Skybond$+$mullite AR coating on flat surfaces, and the epoxy coating on curved surfaces.
}
   \end{figure}

We employ the combination of two-layer AR coating methods, with Skybond $+$ mullite~\cite{Yuki2} on the flat surfaces and with epoxy layers~\cite{Darin,Yuki1} on the curved surfaces as shown in Fig.~\ref{fig:AR}.
The Skybond is polyimide foam made by IST corporation~\cite{IST}.
The thermal sprayed mullite~\cite{Oliver,Toki_thesis} is made by Tocalo corporation~\cite{tocalo}.
The epoxy coating method is with stycast 1090 and 2850FT made by Emerson \& Cuming corporation~\cite{EandC}.
Grooves are made on epoxy surfaces by laser cutting to reduce the thermal stress~\cite{Yuki1,BICEP3}.

%%-----------------------------------------------------------
\section{Focal plane and Detector} 
\begin{figure}
   \begin{center}
   \begin{tabular}{c}
   \includegraphics[height=8cm]{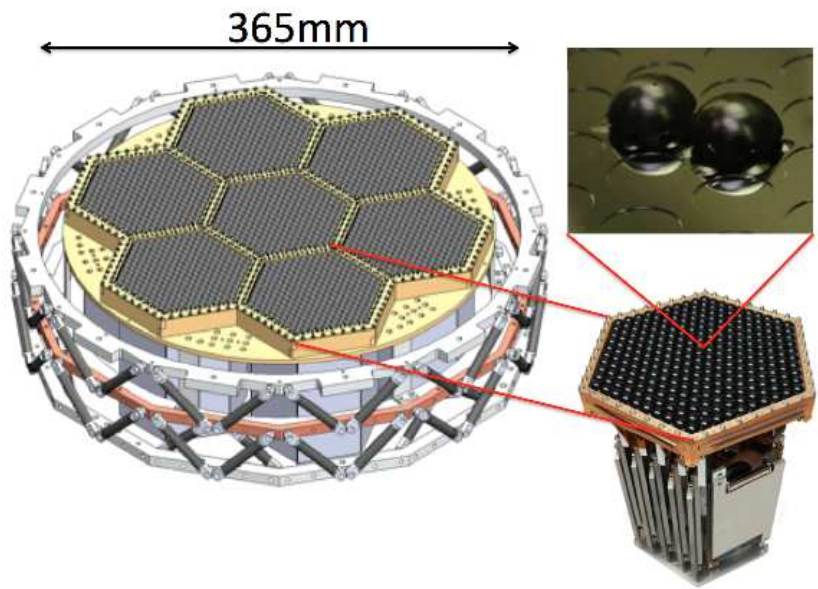}
   \end{tabular}
   \end{center}
   \caption[example] 
%>>>> use \label inside caption to get Fig. number with \ref{}
   { \label{fig:FP} 
Drawing of focal plane. The focal plane consists of 7 wafer modules. Each wafer module has 271 pixels. Each pixel is with a silicon lenslet.}
   \end{figure}

\begin{figure}
   \begin{center}
   \begin{tabular}{c}
   \includegraphics[height=8cm]{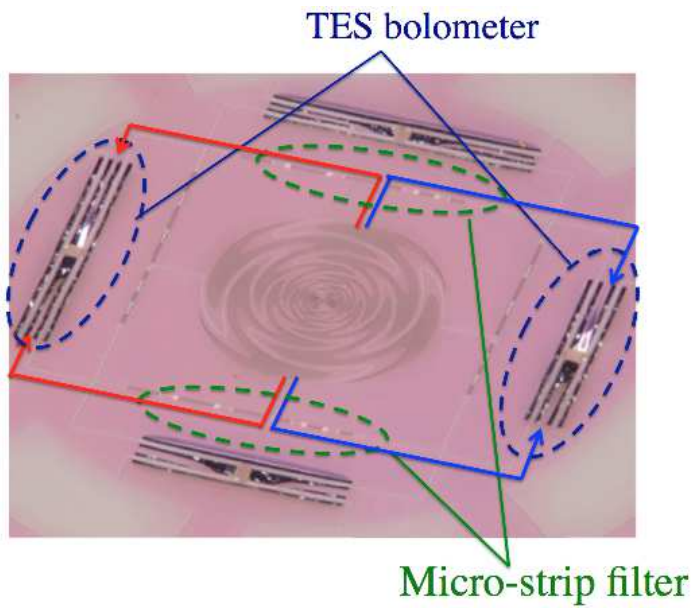}
   \end{tabular}
   \end{center}
   \caption[example] 
%>>>> use \label inside caption to get Fig. number with \ref{}
   { \label{fig:TES} 
Dual-band antenna-coupled Al-Mn TES bolometer. The sinuous antenna is sensitive to 95 and 150 GHz band. The microstrip filter separates the signal between 150 and 95 GHz. The separated signals are detected at TES bolometers.   }
   \end{figure}

\begin{figure}
   \begin{center}
   \begin{tabular}{c}
   \includegraphics[height=4.5cm]{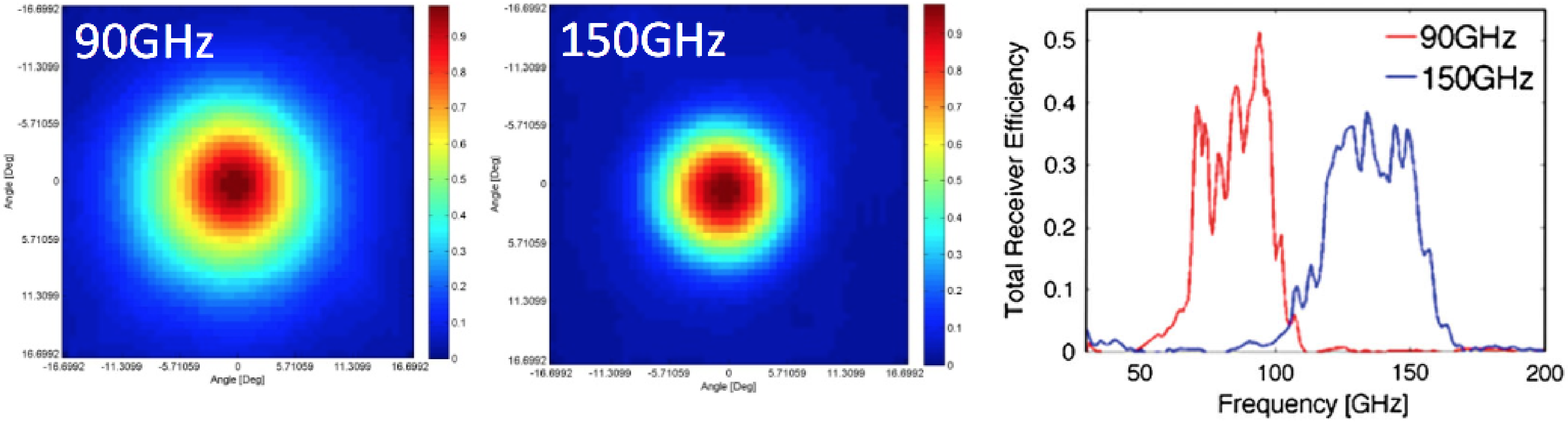}
   \end{tabular}
   \end{center}
   \caption[example] 
%>>>> use \label inside caption to get Fig. number with \ref{}
   { \label{fig:ME} 
Left and middle: The measured beam maps of each detector. Right: the measured bandwidth of microstrip filter. }
   \end{figure}

The focal plane with 365 mm in diameter is placed at the 250 mK stage as shown in Fig.~\ref{fig:FP}. 
Seven hexagonal wafer modules are placed on the focal plane, which consists of detector wafers and LC boards~\cite{Barron:2013xm,Hattori:2013jda}. On each detector wafer, there are 271 pixles with silicon lenslets~\cite{Praween}. The diameter of lenslet is 6.07 mm, whose size is optimized to maximize the array NET. We place the dual-polarization sinuous antennas for broadband detection~\cite{Toki}. The beam map and polarization are shown in Fig.~\ref{fig:ME}. The estimated beam ellipticity at 150 GHz and 95 GHz are less than 1 \%. The microstrip filter separates the signal between 150 and 95 GHz, then Al-Mn TES bolometers detect the signals as shown in Fig.~\ref{fig:TES}. The measured band width of the microstrip filter is shown in Fig.~\ref{fig:ME}. The total number of TES bolometers is 7588.

   
%%-----------------------------------------------------------
\section{Readout} 
\begin{figure}
   \begin{center}
   \begin{tabular}{c}
   \includegraphics[height=6cm]{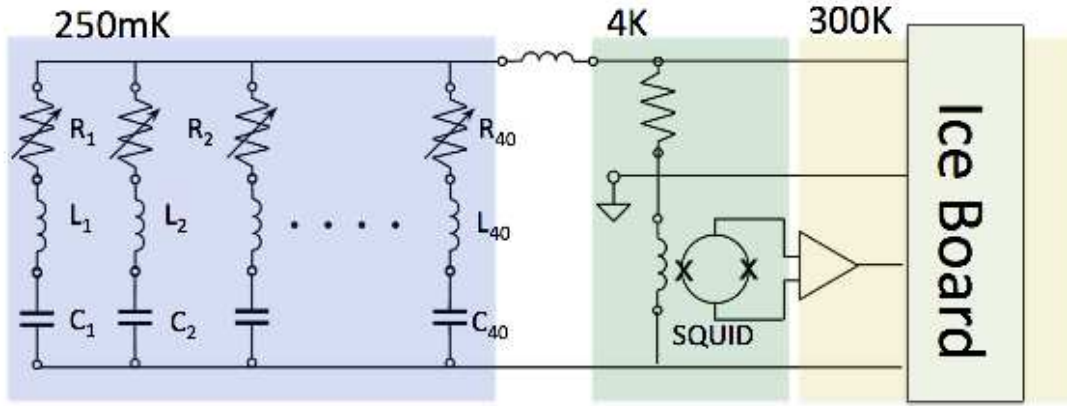}
   \end{tabular}
   \end{center}
   \caption[example] 
%>>>> use \label inside caption to get Fig. number with \ref{}
   { \label{fig:RO} 
The circuit of readout system. TES bolometes and LC chips are placed on 250 mK stage. We read each of the LC resonance peak with SQUID amplifier with superconducting coil. All the read out system is operated with ICE board at room temperature.}
   \end{figure}

The TES bolometers are operated with electro-thermal feedback that keeps the sum of optical and electrical power to be constant~\cite{Hattori:2013jda,Barron:2013xm}.
The resistance of TES bolometer is changed by the incident optical power, which is measured by the superconducting quantum interference devices (SQUIDs).
The SQUID amplifier is operated at 4~K, coupled trough a superconducting input coil.
The readout channels are defined by LC filter with an inductor and a capacitor chip with each bolometer.
Then, these are $60~\mathrm{\mu H}$ inductors and capacitors made by NIST.
The target of multiplexing factor is 40. The readout frequencies are between 1 and 3~MHz. The requirement of electrical cross talk is less than 1 \%. The layout and spacing of LC chips are optimized with minimal cross talk.
All the readout system is controlled by the ICE board~\cite{Amy2014} at room temperature as shown in Fig.~\ref{fig:RO}. The overall expected readout noise is designed to be less than $7~\mathrm{pA \sqrt{Hz}}$. 

%%%%%%%%%%%
\section{Summary} 
PB-2 is an receiver system for performing high-sensitivity observations by placing 7,588 detectors on the focal plane of 365 mm in diameter. 
We have developed and characterized the PB-2 receiver system for performance precision measurements of the B-mode. The PB-2 receiver system will be deployed in Chile in 2017.
%%%%%%%%%%%%%%%%%%%%%%%%%%%%%%%%%%%%%%%%%%%%%%%%%%%%%%%%%%%%%
\acknowledgments     %>>>> equivalent to \section*{ACKNOWLEDGMENTS}       
The POLARBRAR project is funded by the NSF under grant AST-0618398, AST-1212230 and NASA grant NNG06GJ08G.
KEK authors
were supported by MEXT KAKENHI Grant Numbers
JP21111002, JP15H05891, and JSPS KAKENHI Grant Numbers JP13J03626, JP24740182, JP24684017, JP15H03670, JP24111715
and JP26220709. This work was supported by the JSPS Core-to-Core Program. Advanced Research Networks.
The McGill authors acknowledge funding from the Natural Sciences and Engineering Research Council and Canadian Institute for Advanced Research. 
YI was supported by Advanced Research Course in SOKENDAI (The Graduate University for Advanced Studies), and by Academia Sinica under Grants No. CDA-105-M06 in Taiwan.

%%%%%%%%%%%%%%%%%%%%%%%%%%%%%%%%%%%%%%%%%%%%%%%%%%%%%%%%%%%%%
%%%%% References %%%%%

\bibliography{report}   %>>>> bibliography data in report.bib
\bibliographystyle{spiebib}   %>>>> makes bibtex use spiebib.bst

\end{document}